\documentclass[%
 reprint,
 amsmath,amssymb,
 aps,
] {revtex4-2}

\usepackage{graphicx}
\usepackage{dcolumn}
\usepackage{bm}
\usepackage{natbib}

\usepackage{color,soul}
\usepackage[separate-uncertainty=true]{siunitx}
\usepackage{graphicx, wrapfig} 
\usepackage[table]{xcolor}

\usepackage{longtable}
\usepackage{tabularx}
\usepackage{xtab,afterpage}

\usepackage{makecell}
\usepackage{multirow}
\usepackage{enumitem}

\begin{document}

\title{Students' reasoning in choosing measurement instruments in an introductory physics laboratory course \\}

\author{Micol Alemani}
\email{Contact author: alemani@uni-potsdam.de}
\affiliation{Institute of Physics and Astronomy, University of Potsdam, Potsdam 14476, Germany}

\author{Karel Kok}
\affiliation{Physics Education, Department of Physics, Humboldt-Universit\"at zu Berlin\\Newtonstr. 15, Berlin 12489, Germany}

\author{Eva Philippaki}
\affiliation{Department of Physics, Faculty of Natural, Mathematical and Engineering Sciences, King's College London, Strand, London, WC2R 2LS, United Kingdom}

\date{\today}

\begin{abstract}
The aim of this study is to investigate the decisions and reasoning of undergraduate students when choosing simple measurement instruments in an introductory physics laboratory course. For this study, we have developed a questionnaire and implemented it in a pre-/post-test manner to analyze the influence of lab instruction on both students' decisions and reasoning. To characterize students' justifications, we have inductively developed a coding manual that captures the nuances of students' reasoning when choosing an instrument. It shows that students consider different aspects for their decisions, such as data quality, practical and personal considerations.
We have also found that laboratory instruction influenced both students' decisions and justifications, leading to a stronger emphasis on data quality. In fact, after instruction, the majority of students choose the instrument with lower uncertainty and base their justifications mainly on the aim of reducing uncertainties, avoiding systematic effects or mistakes in the instrument reading, and less often than before instruction on personal experience and intuition. 
These findings suggest that dedicating specific laboratory instruction sessions on measurements and data quality, and having students choose between different instrumentation and provide a justification for their decision, can positively impact students' habits in the laboratory and encourage them to base their choices on evidence rather than intuition.

\end{abstract}

\maketitle

\section{\label{sec:Intro}Introduction}

Laboratory courses have always been considered a crucial part of the physics curriculum. While the theory and application of physics are typically taught during lectures, experimental work has the potential to teach students and engage them in the practices used by scientists \cite{Kozminski2014}. This is in fact an important goal of science education. 
In contrast to a somewhat rigid definition of ``the scientific method'', a scientific practice is discussed in the literature as ``a set of regularities of behaviors and social interactions that, although it cannot be accounted for by any set of rules, can be accounted for by an accepted stabilized coherence of reasoning and activities that make sense in light of each other and in light of the practice’s aim. It stabilizes because it successfully addresses critique'' \cite{Ford}.
Most importantly, doing science is characterized by cycles of evaluation and critique of performances according to their goal \cite{Ford}.

An exemplary description of the practices of an experimental physicist can be found in the work of \citeauthor{WiemanPhysToday} \cite{WiemanPhysToday}. They consist of many processes in which a scientist makes decisions after careful evaluations, for example when establishing a research goal and question, defining the criteria for suitable evidence, determining the feasibility of the experiment, selecting materials and apparatus, constructing and testing it,  troubleshooting it, deciding how to analyze data, evaluating results, and presenting them to the scientific community. 
It is suggested that, to learn the processes and practices of science, students must iteratively engage with these extensive scientific evaluation and decision making processes and that this can only happen if students are afforded agency (i.e., they have the opportunity to make decisions to pursue a goal) \cite{PhysRevPhysEducRes.16.010109}. This is similar to the reasoning by \citeauthor{chinn2002} \cite{chinn2002} who suggest that engaging students in authentic inquiry, where students have to make decisions in their experiments, aids them in their scientific reasoning.

One aspect of scientific decision making and reasoning is to have students choose their own measurement instruments. 
However, in most laboratory courses, the instrumentation that students use has most likely been pre-selected and pre-otpimized by the instructors, so there is no room for practicing agency in this regard. In contrast, in authentic situations and as discussed above \cite{WiemanPhysToday, chinn2002}, scientists must often evaluate and decide which instrument is better to use depending on the ultimate experimental goal and there can be more than one instrument appropriate for that goal. 

In this study we look into students' reasoning when choosing between measurement instruments. We start by discussing research-based laboratory courses and how they support agency for students. We continue by describing previous work that has been done on how students choose measurement instruments. We end our theory by stating the aims and research questions of this work.

\subsection{Research-based laboratory courses}

In the literature about laboratory courses, we find on the one hand, the traditional ``cookbook'' courses with the goal of reinforcing physics concepts. In these courses, students typically follow a list of instructions to confirm theory and work on set-ups pre-designed and pre-optimized by their instructors. These courses have been recurrently criticized over the years \cite{May2023} and are characterized by low levels of student agency \cite{PhysRevPhysEducRes.16.010109}. Recently, they have also been shown to be ineffective for students' development of content knowledge \cite{holmes2017value}, for the acquisition of experimental abilities \cite{karelina2007acting}, for a deep understanding of measurement uncertainties \cite{PhysRevSTPER.4.010108,PhysRevPhysEducRes.16.020160}, and for developing "expert-like" views of experimental physics \cite{wilcox2017developing}. 

On the other hand, new types of laboratory curricula have recently been implemented that allow students with more agency \cite{Kalender_Holmes} and that have been shown to be effective in developing students' experimental skills \cite{Etk_Karelina, etkina2006using, Holmes, Zwickl2013a, AlemaniLabReconstr}, a deep understanding of measurement \cite{buffler2008teaching,kung2005teaching, Liu2025}, and for fostering "expert-like" views of experimental physics \cite{wilcox2017developing, Liu2025}. We refer to these courses as research-based courses. Upon these research-based courses, some address the need in our modern world to develop students critical thinking skills \cite{holmes2015teaching, walsh2019quantifying, Smith_Holmes}.
Those skills are in fact useful for preparing the new generation of physicists in making decisions based on data and self-reflection in complex situations where information can be false.

As discussed in the work of \citeauthor{PhysRevPhysEducRes.16.010109} \cite{PhysRevPhysEducRes.16.010109}, supporting students in decision making processes associated with scientific practices requires some scaffolding. For example proving guiding questions to make all important decisions explicit or prompting students to reflect and provide justifications for their actions. This structure can be reduced over time, while students acquire habits of mind and expectations of behavior to adopt scientific practices.

\subsection{\label{sec:LitReview} Literature review on students' reasoning when choosing instrumentation}

Although there are several studies about students' understanding of measurement uncertainties, data quality, and validity \cite{PhysRevPhysEducRes.14.010121,allie1998first,PhysRevPhysEducRes.15.010103,PhysRevSTPER.4.010108}, topics related to students' evaluation of measurement set-ups and devices, research on how students choose instrumentation remains very limited.

A recent study has identified misconceptions about the way students evaluate instrumental precision, and found that students sometimes struggle to identify the precision of a measurement in relation to the instrument used to make that measurement. In addition, students often showed to have more confidence in digital scales than the measurement uncertainty would suggest \cite{GeschwindMesUnc}. 

From a instructors' point of view, the study by \citeauthor{Irwansyah_Slamet_Ramdhani_2018} \cite{Irwansyah_Slamet_Ramdhani_2018} in chemistry education shows that key factors relating to laboratory equipment selection are: safety, improvement of practitioner’s understanding, ease of use, the accuracy level of equipment, and cost. Although these aspects are important, it does not shed light on students' reasoning.

Another recent study by \citeauthor{logman2025} \cite{logman2025} examined the decision-making process of undergraduate physics students halfway through their first year lab course, when deciding between two set-ups for measuring electrical signals, (i) a set-up consisting of a function generator and an oscilloscope and (ii) a bench combining these two instruments in one. 
Students could express their preference and use their chosen set of equipment for their experiment, without being informed by the teaching team about any specific technical advantages of the equipment.
The authors used the \citeauthor{Svenson}'s four levels of decision-making as theoretical framework \cite{Svenson} and identified which characteristics of a lab course foster higher levels of decision-making when choosing experimental instruments. These four levels are:
\begin{description}
    \item[Level 1] includes quick and automatic decisions, like choosing apparatus without thinking. Students stated no reason for their preference or stated that they had no preference.
    \item[Level 2] is characterized by students noticing the set of alternative apparatus yet going for the most appealing choice among the available options. Students gave an intuitive reason for choosing their equipment without giving any technical facts to justify their choice.
    \item[Level 3] is reached when choosing one alternative over the other is based on facts, not stereotypes or emotions. Students used technical facts about the two alternatives to substantiate their decision.
    \item[Level 4] The decision maker faces a problem in which new alternatives to choose from are created. Students asked for other equipment than the available two options.
\end{description}

The study found that \SI{63}{\percent} of students would choose an instrument based on ease of use (Level 2), \SI{20}{\percent} due to technical reasons (Level 3) while \SI{17}{\percent} would give no specific reason (Level 1). None of the students asked for other equipment than the one they had experienced (Level 4)\cite{logman2025}.

The low levels of decision making are in accordance with the findings by \citeauthor{PhysRevPhysEducRes.14.010121} \cite{PhysRevPhysEducRes.14.010121} comparing the views about validity of experimental results of introductory students and PhD students’. It was shown that when undergraduate students aim to improve the quality of results of an experiment, they primarily focus on representational roles (e.g., describing imperfections of the experiment and apparatus), rather than uncertainty analysis and experimental methodology. On the other hand, advanced students focused on the inferential roles of uncertainty analysis (e.g., quantifying reliability, making comparisons, and guiding refinements) \cite{PhysRevPhysEducRes.14.010121}.

\subsection{Aims of our study}

Our study seeks to further investigate this under-researched field of students' decision-making processes when choosing measurement instruments.

In fact the study by \citeauthor{logman2025} compares students' reasoning with the four levels of decision-making by \citeauthor{Svenson} which is rooted in psychology and human cognition. Our study aims to shed more light onto students' reasoning when choosing a measurement instrument by taking an inductive approach to look at students' reasoning. We will do this by answering the following research questions:

\begin{itemize}
    \item[\textbf{Q1:}] What are the decisions and justifications that undergraduate students give when choosing between two measurement instruments, at the start of their introductory physics lab course?
    \item[\textbf{Q2:}] How do students' decisions and justifications change after receiving lab course instruction?
\end{itemize}

\section{\label{sec:Methods}Methods}

The study was conducted at the University of Potsdam from the summer semester of 2024 until the end of the winter semester 2024-2025. It was approved by the ethical commission of the University of Potsdam (proposal number 99/2023).

Data were collected using a pre--post questionnaire. 
The pre-test was conducted prior to instruction, during the initial course session that provided a general introduction and safety training. The post-test was administered subsequently, after students had completed two laboratory sessions focused on the introduction to measurement uncertainties, a session (only in the courses for physicists and pre-service teachers) about laboratory notebooks, and a session using the instrumentation of the questionnaire.

Participation was voluntary and students completed the questionnaire digitally during lab class time. The completion took about 10--15 minutes.

\subsection{\label{subsec:CourseContext} Course context and study participants}

In total, 231 undergraduate students participated in our study. The study was conducted in the introductory courses for physicists, pre-service physics teachers, chemists, geologists, geoecologists, biologists, and nutritionists. 
The introductory physics lab course for physicists and pre-service physics teachers is the first part of a sequence of lab courses and takes place during the first semester of study. For the non-physics majors the course is not part of a sequence and takes place in further semesters. For the biologists this is happening at the end of the first semester in year one, for the geologists and chemists during the second semester in year one, while for the geoecologists during their first semester in year three (or fifth semester).

The laboratory courses at the University of Potsdam were reconstructed by one of the authors (M.A.) starting in 2016 using a research based approach. Goal of the reconstruction was to prepare students for independent experimental work, offering them authentic experiences that foster experimental skills acquisition and "expert-like" attitudes towards experimental physics, as described in previous works \cite{AlemaniLabReconstr, TeichmannGE-CLASS, Alemani2023}. For this study, it is important to mention that students work in the lab typically in groups of three, and that the course starts with an introduction to measurement uncertainties (lasting two sessions) and (only for physicists and pre-service teachers) a session on laboratory notebooks, followed (for all students) by a practical laboratory session in which the students practice using the simple instrumentation of the questionnaire. 

Specific learning objectives regarding data quality for the course are the following:

\begin{itemize} [label=-]
 \item Students recognize that every measurement has an associated measurement uncertainty. As a consequence of this, experimental results are given complete of their measurement uncertainties
 \item Students can distinguish between Type A and Type B determination of measurement uncertainties
 \item Students can recognize the difference between standard deviation and standard deviation of the mean
 \item Students are able to identify sources of experimental uncertainty for direct measurements 
 \item Students are able to distinguish between measurements uncertainties and systematic effects and between precise and accurate measurements
 \item Students are able to identify possible sources for systematic effects
 \item Students can calculate the mean, standard deviation and standard deviation of the mean
 \item Students are able to estimate measurement uncertainties for direct (including combining different types of uncertainties) and indirect measurements and describe how this estimation was performed in a clear way
\end{itemize}

Note that the two sessions on measurement uncertainties are implemented using active learning methods. In particular, before both of these two lab sessions, students had to look at some videos and read some documents that were provided through a learning management system. In class, students had the opportunity to ask questions about the content of the videos and documents and worked on problem exercises discussing them in groups. Further exercises are given to the groups as homework and feedback is provided.\newline
\indent The introduction on laboratory notebooks for physicists and pre-service teachers lasts one lab session, and it is inspired by the works by  J. T. Stanley, J. R. Hoehn, and H. J. Lewandowski \cite{Stanley, Stanley2, Hoehn}. As a preparation for this lab session, students have to watch a video about laboratory notebooks. The in-class activity consists of an initial group discussion on what laboratory notebooks are (including the difference with laboratory reports), why they are used daily by physicists (i.e., their purpose) , what type of information is written in them, what is important to consider when writing them, and what are the difficulties when writing them. The discussion is followed by three exercises. The first consists of reviewing a (student) lab book, discussing what aspects have been successfully realized and what aspects could be improved. The second exercise consists of reviewing a (students) lab project report and identifying information that is relevant for a lab book but not for a report. The third exercise consists of watching a video of a student who was given the task of measuring the density of a cherry tomato and a cucumber with two different methods and write a laboratory book for that experiment. While students receive feedback on the first two exercises directly in class during a class discussion, they receive specific (for each student group) feedback on the submitted laboratory book of the third exercise.\newline
\indent The introduction to laboratory notebooks for non-physics majors is integrated into the course without a dedicated session with activities. This is due to the much shorter duration of the course. However, students receive the video about lab books to watch at home and written material to prepare before the first practical lab session, which begins with a class discussion about lab books. \\
After the two sessions about measurement uncertainty (and the session about lab books), there is a session dedicated to perform simple (length) measurements using the instrumentation that are present in the questionnaire (plus an analog caliper). Before coming to the lab, students have to look at some videos and use online applets to become familiar with reading Vernier scales. They also have to complete a quiz with questions about concepts of precision and accuracy and demonstrate their ability in reading the Vernier scales of micrometers and calipers. In the lab, we ask students to measure the dimensions of some objects and reflect on their measurements (e.g., precision, accuracy).

\subsection{\label{subsec:Questionnaire} Questionnaire}

The questionnaire comprises four items. For each item, students are asked to select, from a pair of instruments, the one they would use to measure the diameter of a metallic rod approximately \SI{1.5}{cm} in size, or, in the case of item 4, the length of a laboratory table. In the multiple-choice section, students may choose one of the two instruments or indicate no preference by selecting the option “Either of the two.”

The instrument pairs used in each item are listed in Table~\ref{tab:1}. 
They include instruments that are commonly used in physics laboratory courses. These were: (1) a digital caliper, (2) an analog micrometer, (3) an analog dial caliper, (4) a digital laser distance meter, and (5) an analog measuring tape. A reference table with the technical specifications of all instruments'—names, measurement ranges, resolutions, and accuracies was also provided to students.

\begin{table}[]
    \caption{Overview of the instrumentation included in the student questionnaire.}
    \label{tab:1}
    \begin{tabular}{cc}
        \hline\hline
        \hspace{3mm}Item Number & \hspace{4mm}Choice of Instrument\hspace{4mm}\\
        \hline
        \multirow{2}{*}{1} & \makecell[c]{Digital Caliper}\\ 
         & \makecell[c]{Analog Micrometer} \\ 
         \hline
        \multirow{2}{*}{2} & \makecell[c]{Digital Caliper} \\ 
         & \makecell[c]{Analog Dial Caliper} \\ 
         \hline
        \multirow{2}{*}{3} & \makecell[c]{Analog Micrometer} \\ 
         & \makecell[c]{Analog Dial Caliper} \\ 
         \hline
        \multirow{2}{*}{4} & \makecell[c]{Digital Laser Distance Meter} \\ 
         & \makecell[c]{Analog Measurement Tape} \\ 
        \hline\hline
    \end{tabular}
\end{table}
In addition to selecting an instrument, students were invited to justify their choice in an open-ended response field. 
Such a design of the questionnaire with a multiple choice part and an open-ended part is motivated by our desire to give students the opportunity to make a decision and, at the same time, to understand why that decision is made. In this way, it is possible to investigate students' reasoning when choosing instrumentation and to analyze not only why students choose a certain instrument but also whether their choice of instrument is consistent with their justification. At the same time, this also helps to detect misalignments in decisions with justifications \cite{kok2024}.

We note that the students could interpret the response option “Either of the two” in different ways. On the one hand, they could choose it to indicate uncertainty because, for example, they don't know any of the two instruments. On the other hand, they could choose this option when they considered both instruments appropriate for the measurement, for example because they see pro and contra for both instruments. In the data analysis (see section \ref{subsec:StJustifications}) we will analyze these possibilities in detail.

In the questionnaire, the only detail provided about the measurement is (for items 1 to 3) that the goal of the measurement is to measure the diameter of a metallic rod of about \SI{1.5}{cm} in diameter. However, we did not provide further details about the context or the goal of the measurement. For example stating that “the goal of the measurement is to obtain a measurement as precise as possible because the metallic rod must fit into a cylindric hole with a diameter of \SI{1.52+-0.55}{cm}”.
This design choice of the questionnaire could be considered as missing important information, since in authentic situations the decision for an instrument is strictly related to the specific context and goal of the measurement. For example, if the goal of a measurement is to measure something quickly and high precision is not required, it would be a waste of time to use the most precise instrument if its usage would require higher time investment. On the other hand, if an experimentalist wants to measure something as precise as possible, they would choose the instrument with lower uncertainty. Moreover, in authentic situations, instrumental choices can also be based on familiarity, experience, time, budget constraints or just availability of instruments in the laboratory. \newline
\indent However, in our questionnaire, our decision to put no further context details about the goal of the measurement was done to prevent the introduction of external “expert” biases that could influence students’ thinking. \newline
\indent Moreover, we want to get as wide a range as possible for students' considerations in their decision for a specific measurement instrument. Putting ``constraints'' on the goal of the experiment would narrow this down.
Therefore, we do not even want to assume that students understand the relationship between instrumental choice and measurement goal, but aim to explore this aspect in our study. We provided students the option "Either of the two" to allow them to express the dependency on the situation and goal of the measurement.\newline
\indent Our selection of the specific instruments in the questionnaire is due to their shared, straightforward purpose (i.e., measuring distance) and by their simplicity. Except for the laser distance meter, all other instruments included in the questionnaire perform direct length measurements by comparison with a scale. In contrast, the laser distance meter functions on a more sophisticated principle, as it calculates the distance indirectly by measuring the time taken for an emitted light signal to travel to a reflective surface and back to the sensor, assuming specific air conditions (e.g., temperature) when using the speed of light for the calculation. 
Since the laboratory course part here described had not yet introduced exercises involving the modeling of complex measurement instruments or systems (topic of the following laboratory course meetings/sequences) and since the working mechanism of the laser meter is hidden from the user but the measurement result is directly expressed in units of meters, we assume that students do not necessarily elaborate a model of this instrument.

\subsection{\label{subsec:AnalysisJust} Analysis of justifications}

To answer research question Q1, we have analyzed students' justifications for their decisions. This was done in an inductive manner through an iterative process of reading, identifying categories and subcategories of codes, phrasing coding rules, double coding, and discussion, similar to the process of \citeauthor{kok2023c} \cite{kok2023c}.

With the coding manual that emerged from this process, a single justification can have more than one code. This can be troublesome during the inter-rater reliability calculation, where one coder could have assigned more codes than the other. In these instances, a place-holder code was assigned to the coder who had less codes so that the total number of codes assigned was the same.

\section{\label{sec:Results}Results}
First, we will present the coding manual that was used to analyze students' justifications. This is followed by the results of the questionnaire, starting with the decisions after which we present the analysis of the justifications using our coding manual.

In the analysis of the decisions and justification codes, a Pearson’s chi-squared test for homogeneity was used to look at differences between distributions. This tests whether two or more distributions share a common distribution. The effect size of the difference was calculated using Cohen’s $w$. In cases where the distribution of codes violated the Pearson’s chi-square test assumptions (no cells with zero’s, and \SI{80}{\percent} of the cells have to have a value $>$5), a Pearson’s chi-squared test of independence with simulated $p$-values (10,000 replicates) was calculated.

\subsection{\label{subsec:CodeMan}The coding manual}

After five rounds of coding and rephrasing, we have identified three main categories of justifications, supplemented by a category for \emph{uncodable} responses (labeled as \textit{unc.}). These three categories are for justifications that are based on: \emph{data quality} (labeled as \textit{data}), \emph{the experimental process} (labeled as \textit{exp.}), and \emph{personal considerations} (labeled as \textit{pers.}). Each of these categories contains one or more sub-categories that give a fine-grained insight into students' reasoning. An overview of each sub-category can be found in Tab.~\ref{tab:coding_manual}. The complete coding manual with examples is presented in the Appendix, Tab.~\ref{tab:coding_manual_full}). 
Note that although the codes for the sub-categories contain numbers, the codes do not have a hierarchical order.

\begin{table*}
\caption{\label{tab:coding_manual} Description of the coding manual used to characterize students' reasoning. }

\begin{ruledtabular}
\begin{tabularx}{\textwidth}{c p{16cm}}

 Code&Code description\\ \hline
 data.1 & The quality of the data is considered for the decision of an instrument. In particular the precision and/or accuracy of the instrument and/or the correctness and/or the overall uncertainty of the measurement results are taken into account.\\
 \hline
 data.2 & Avoiding mistakes, systematic effects, or measurement uncertainties by reading off the instrument's scale and it's effect on the data are explicitly considered for the decision of an instrument.\\
  \hline
 exp.1 & Practical (non-personal) considerations specific to the conditions of the experiment, the procedure of taking measurements, or the object to be measured are considered for the decision of an instrument.\\
  \hline
 pers.1 &  Personal (in)experience in using or knowledge of an instrument is considered for the decision of an instrument.\\
  \hline
 pers.2 & Practical personal considerations specific to the instrument usage such as ease of use or speed are considered for the decision of an instrument. \\
  \hline
 pers.3 & Personal preference towards an instrument is indicated for the decision of an instrument, this can be based on gut-feeling or personal bias.\\
  \hline
 pers.4 & Fun and/or enjoyment in using an instrument are considered.\\
  \hline
 unc.1 & The justification is not codable because it is either not clear or it does not relate to the decision for either of the instruments.\\
\end{tabularx}
\end{ruledtabular}
\end{table*}

In order to clarify here how the sub-codes described in Tab.~\ref{tab:coding_manual} were assigned, we provide in the following representative examples of students' justifications and discuss the assignment of the respective sub-codes.
A good example for sub-code \textit{data.1} is the following:

\begin{quote}
    Da der digitale Messschieber eine geringere Messunsicherheit hat. [GER]\\
    Because the digital caliper has a smaller measurement uncertainty. [ENG, authors]
\end{quote}

This justification refers to the measurement uncertainty of the caliper being lower than the other instrument. Since a smaller measurement uncertainty will increase the quality of the data. Expressions like this are coded \textit{data.1}.

The difference between codes \textit{data.2} and \textit{pers.2} can be subtle. Some students mention that the reading can be taken easily:

\begin{quote}
    Der erste Messschieber ist mir bereits vertraut und ich weiß genau was ich tun muss. Zudem ermöglicht die digitale Anzeige eine einfacheres Ablesen. [GER]\\
    I am already familiar with the first caliper and know exactly what to do. In addition, the digital reading makes it easier to read. [ENG, authors]
\end{quote}

The last part of this justification mentions the ease of taking a reading. However, it does not mention the influence on the data quality (e.g., the accuracy or overall uncertainty). Therefore, it is not coded \textit{data.2} but rather \textit{pers.2}. Also, the justification mentions personal experience with the instrument so the justification has two codes:\ \textit{pers.1} and \textit{pers.2}. Only when students mention the effect that taking a reading has, justifications can be coded \textit{data.2}:

\begin{quote}
    Weil da die Messunsicherheit des Ablesefehlers wegfällt und die Messunsicherheit des Gerätes beträgt bei beiden 0,1g [GER]\\
    Because the measurement uncertainty of the reading error is eliminated and the measurement uncertainty of the device is 0.1 g for both. [ENG, authors]
\end{quote}

Despite the fact that this student made a mistake in the units, this justification is also coded with \textit{data.1} since it makes reference to the uncertainty of the instruments (probably, falsely, considering the absence of a reading uncertainty of the instrument).

This false reasoning about digital instruments not having reading uncertainty was found also in further examples.

Justifications that are coded \textit{exp.1} refer to practical considerations specific to the experiment or its procedures. This can be references to e.g., the range of the instrument and the expected value of the measurement, the specific object that is being measured, or practical procedures:

\begin{quote}
    Laser benötigt ein Ende an dem man die reflektion dann erkennen kann [GER]\\
    The laser requires an end point on which the reflection can be seen [ENG, authors]
\end{quote}

We note that within the data-category we do not distinguish between measurement uncertainties, systematic effects, accuracy, or precision. This is due to the occasional inconsistent and interchangeable use of terminology by students when discussing measurement uncertainties (i.e., using the word ``Fehler'' [error] to refer to ``Unsicherheit'' [uncertainty]), which made it difficult to reliably distinguish between these concepts. This issue will be explored in more detail in the following section \ref{subsec:StDecisions}.

To measure the reliability of the coding manual, we have randomly pulled 50 justifications from the data set and two authors double-coded these justifications. The inter-rater reliability in form of Cohen's Kappa showed good agreement between the coders, $\kappa=0.76$ \cite{LandisKoch}. Afterwards, the remaining justifications were coded by one author.

\subsection{\label{subsec:StDecisions}Students' decisions}

\begin{figure}[t!]
\includegraphics[width=.6\linewidth]{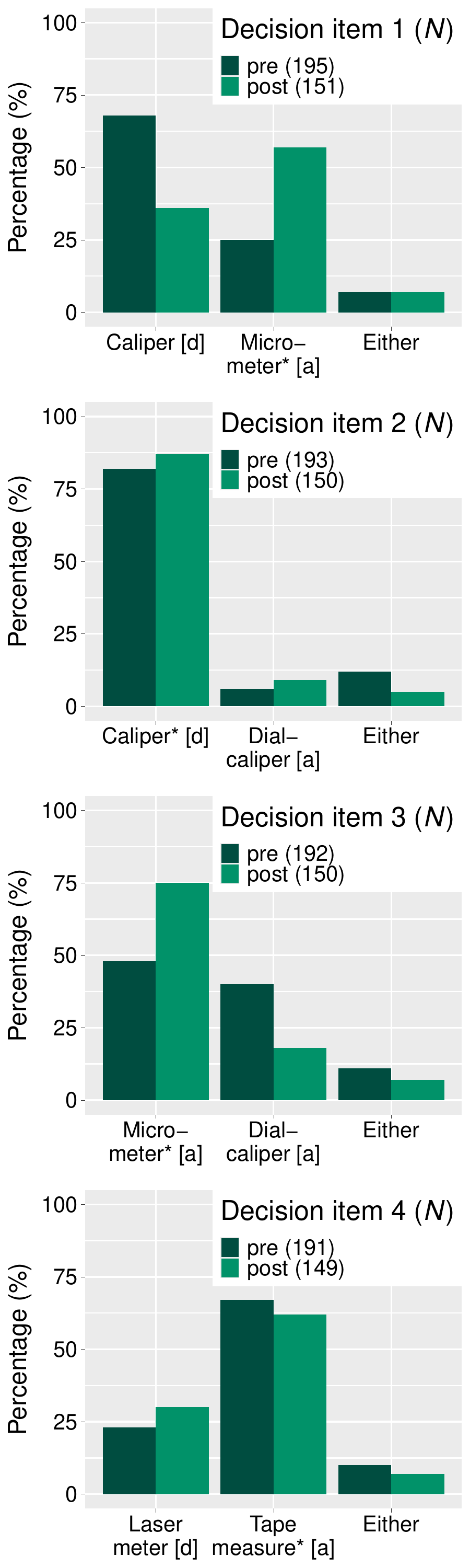}
\caption{Students' answers to the multiple choice part of the questionnaire. On the $x$-axis are indicated the three possible answers for each item. Top to bottom graphs show respectively items 1 to 4. Pre- and post-test data are indicated with dark and light green. The number $N$ in parenthesis indicates how many students chose each particular answer. The type of instrument (digital or analog) is indicated for each instrument as $[d]$ and $[a]$ respectively.}
\label{fig:decisions}
\end{figure}

We analyzed which instruments students choose, by looking at students' answers to the multiple choice questions. We note that for each multiple choice question, students can express their preference for a specific instrument (between the two possible instruments) or express no particular preference (choice "Either of the two").
The histograms of the students' decisions are shown in Fig.\ \ref{fig:decisions} for each item separately. 
The percentage of answers shown on the ordinate takes into account only given answers, i.e., we did not count when a student did not provide an answer. For each item, in the figure we indicate as $N$ the number of answers given for each item. In the following, we refer to the percentage of students as the percentage of students who provided an answer.
The results for the pre-questionnaire are shown in dark green, while results of the post-questionnaire are reported in light green. 
We have indicated in the figure the instrumentation with lower measurement uncertainty with an asterisk. We considered for the determination of the uncertainty the combination of the one related to the reading scale and the one related to the instrumental accuracy.

For item 1, in the pre-questionnaire the majority of students chose the digital caliper \SI{68}{\percent}, while only \SI{25}{\percent} selected the analog micrometer or the choice "Either of the two", \SI{7}{\percent}. The situation changes in the post-questionnaire in which \SI{36}{\percent} of the students opted for the caliper, while \SI{57}{\percent} for the micrometer, which is the instrument with lower uncertainty. For this item, we found a statistical significant difference between the distributions of students' answers in the pre- and post-test [$\chi^2(2)=39.55$, $p<0.001$]. This difference has a medium size effect ($w=0.34$). We note that the difference in the distribution is entirely due to a major change in the choices between the two instruments and not in a change of the choice "Either of the two", which occurs with the same frequency in the two questionnaires.

For item 2 students unanimously prefer the digital caliper in both the pre- (\SI{82}{\percent}) and post-questionnaires (\SI{87}{\percent}), which is the instrument with the lower uncertainty.  In fact, less than \SI{10}{\percent} chose the analog caliper (pre and post). The small difference in the distributions of students' answers observable by comparing the pre- and post-answers is not statistically significant ($p>0.001$). 

In contrast, for item 3 students' choices in the pre-test are almost equally divided between the two instrument choices (\SI{48}{\percent} for the analog micrometer and \SI{40}{\percent} for the analog caliper). In the post-test we found a statistically different distribution [$\chi^2(2)=24.86$, $p<0.001$] with small to medium size effect ($w=0.27$) with respect to the pre-test. In fact, in the post-test, students' choices are strongly shifting from the analog caliper (\SI{18}{\percent}) to the analog micrometer (\SI{75}{\percent}), which has the lower uncertainty between the two instruments.

On the last item of the questionnaire, i.e., item 4, the distribution of students' answers is only slightly changing between the pre- and post-test and the difference is not statistically significant. The majority of the students chose the analog measurement tape (\SI{67}{\percent} in the pre-test and \SI{62}{\percent} in the post), which is the instrument with lower uncertainty. Fewer students chose the laser distance meter in the pre- and post-test (\SI{23}{\percent} and \SI{30}{\percent} respectively).

A detailed analysis of the justifications where students chose the answer "Either of the two" (see Table  \ref{tabEitheroftheTwo}) shows that in the majority of cases, students chose the answer "Either of the two" when they think both instruments can be used (either because they are equivalent or because they have pro and contra) and not because they do not know what to use. For this analysis we considered the following options: (i) students do not give a justification; (ii) students consider both instruments appropriate for the measurement because are considered equivalent, or it depends on the situation/goal of the measurement or because they have pro and contra; (iii) students do not know what to decide because for example they do not know both instruments; (iv) students' justification is uncodable because we cannot clearly assign students' reasoning to one of the option (ii) or (iii).

\begin{table}
    \label{tabEitheroftheTwo}
    \caption{Analysis of the answers "Either of the two"}
    \begin{ruledtabular}
    \begin{tabular}{llllr}
   
    Item Number & 1 & 2 & 3 & 4 \\ 
    \hline
        "Both instruments are equivalent"&  15&  22& 15 & 19\\ 
        "I don't know"&  3&  4& 8& 1\\ 
        No justification &  6&  3& 9& 9\\ 
       Uncodable justification &  1&  1& 1& 1\\ 
      \end{tabular}
\end{ruledtabular}
\end{table}

\subsection{\label{subsec:StJustifications}Students' justifications}

\begin{figure*}[bt]
\includegraphics[width=0.9\linewidth]{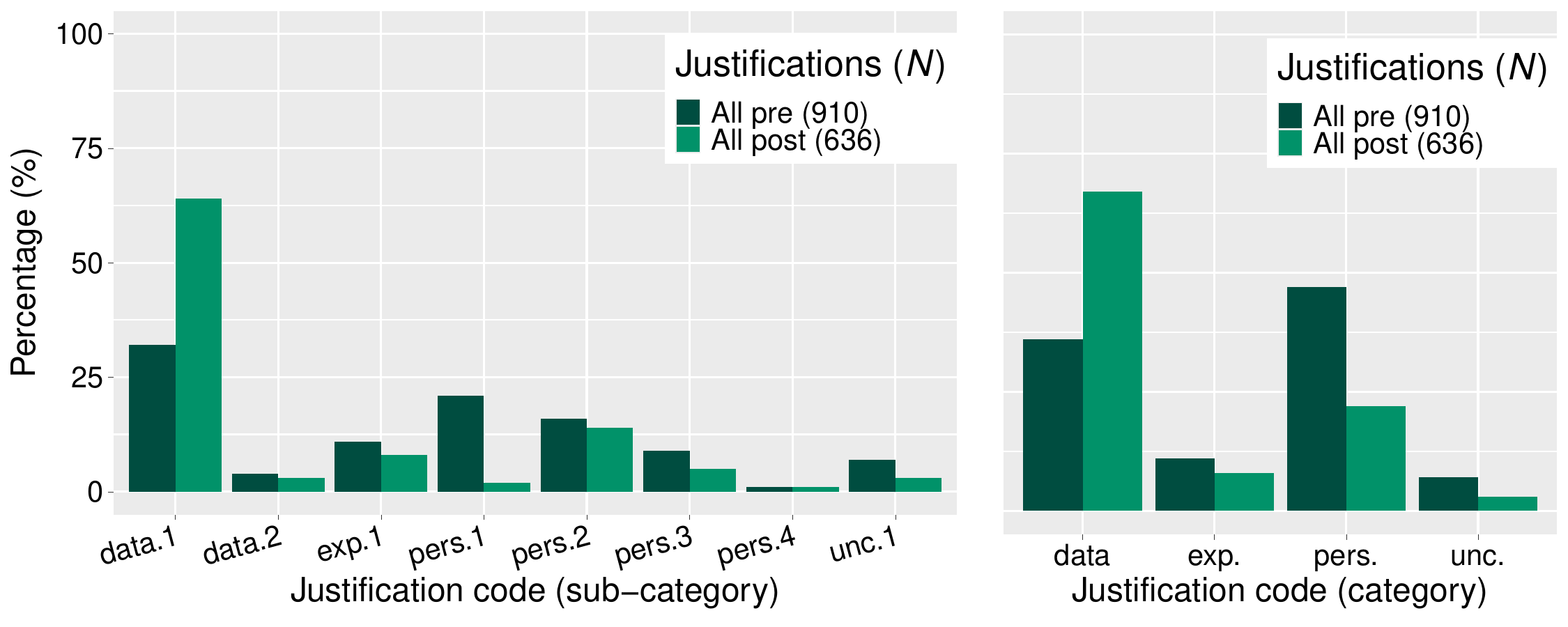}
\caption{Percentage of the students' justifications in the different coding sub-categories (left) and categories (right) considering all items and all kind of instruments. Pre-test data are shown in dark green and post-test data are shown in light green. (Left) On the $x$-axis are reported the coding sub-categories as described in Tab. \ref{tab:coding_manual}. (Right) The coding categories indicated on the $x$-axis are, from left to right, \textit{data}, \textit{exp.}, \textit{pers.}, and \textit{unc.}. The numbers $N$ indicated in the parenthesis indicates the total number of codes assigned in the pre- and post-test respectively.}
\label{fig:justifications_all}
\end{figure*}

In order to investigate students' justifications used to motivate their choices of instruments in the questionnaire, we have coded students' answers to the free-response questions, using the full eightfold coding manual, including the sub-categories, as presented in Table \ref{tab:coding_manual} and in the appendix (see Table \ref{tab:coding_manual_full}). The result of this coding, considering all items and all instruments, is shown on the left side of Fig.\ \ref{fig:justifications_all}. In this figure, the results of the pre-test are shown in dark green, while post-test results are in light green. 

Since a single answer to a questionnaire question can be categorized by more than one code, the percentage of codes reported on the ordinate of the graph does not indicate the percentage of answers or the percentage of students. Rather, it indicates which percentage of the total assigned codes falls into the different coding categories. The sum of the percentages of each category is \SI{100}{\percent} and the total number of codes assigned is indicated as $N$. This percentage can be used to investigate the distribution of students' reasoning as a whole. 

We found that before instruction, students' reasoning is distributed in order of occurrence between the sub-categories \textit{data.1}, \textit{pers.1}, \textit{pers.2}, \textit{exp.1}, \textit{pers.3}, \textit{unc.1}, \textit{data.2} and \textit{pers.4}. However, apart from the sub-categories \textit{data.1}, \textit{pers.1} and \textit{pers.2}, the other sub-categories are not well populated (they stay below \SI{10}{\percent}). In the post-test, the distribution changes and almost all justifications are coded \textit{data.1}. We observe a strong reduction of the sub-categories \textit{pers.1} and \textit{pers.3}.
Even if the population of all eight categories does not allow us to perform statistical analysis (because the occurrences in several of the eight categories are very low), we believe that this extended coding manual allows us to capture all characteristics of students reasoning into details.
However, for further statistical analysis, we considered only the categories of the fourfold coding manual. We refer to the four-fold coding manual as the coding manual "per category". The results of the analysis of students justifications using this coding manual per-category are shown on the right side of Fig.\ \ref{fig:justifications_all}.

We found that before instruction, the vast majority of justifications fall into the category \textit{pers.}\ (\SI{47}{\percent}). The category \textit{data} occurs, on the other hand, in \SI{36}{\percent} of the cases. The categories \textit{exp.}\ and \textit{unc.}\ are much less frequent (respectively \SI{11}{\percent} and \SI{7}{\percent}).

After instruction, students mainly use justifications that fall into the category \textit{data} (\SI{67}{\percent}), i.e., related to data quality, while the category \textit{pers.} becomes less frequent (\SI{22}{\percent}). As discussed above by looking at the sub-categories, this decrease in occurrence arises by a decrease of the subcategories \textit{pers.1} and \textit{pers.3}. This result can be interpreted as students becoming more familiar with the instrumentation during the laboratory course.  Also in the post-test, the categories \textit{exp.}\ (\SI{8}{\percent}) and \textit{unc.}\ (\SI{3}{\percent}) show up less frequently.  For these two categories, a slight decrease in occurrence is observable. 
The difference in the codes distributions per-category in the pre- and post-test is found to be statistically significant [$\chi^2(3)=153.41$, $p<0.001$] and with a small size effect ($w=0.19$). This shows that instruction has an impact on students' reasoning. After instruction, students mostly use data quality aspects in their reasoning and less often aspects based on personal considerations like experience or personal biases.

The result is striking, as despite the absence of explicit references to data quality in the formulation of the questionnaire items, the instruction emphasizing the evaluation of data quality led students to independently apply this consideration as a criterion in their responses.
Unfortunately, as discussed above, we cannot distinguish in the category \textit{data} between the aim of reducing systematic effects and the aim of reducing uncertainties, therefore the interpretation of this focus on data quality is not simple. On the one hand, it is positive that students recognize after the lab course the importance of avoiding having systematic effects in any measurement. On the other hand, as we discussed before, we note that it is not worth for every measurement to aim for the highest possible precision, as one should instead always aim for the needed precision. 
We consider this an important aspect that needs to be clarified in a lab course when teaching about measurements. 

\begin{figure*}[bt]
\includegraphics[width=1\linewidth]{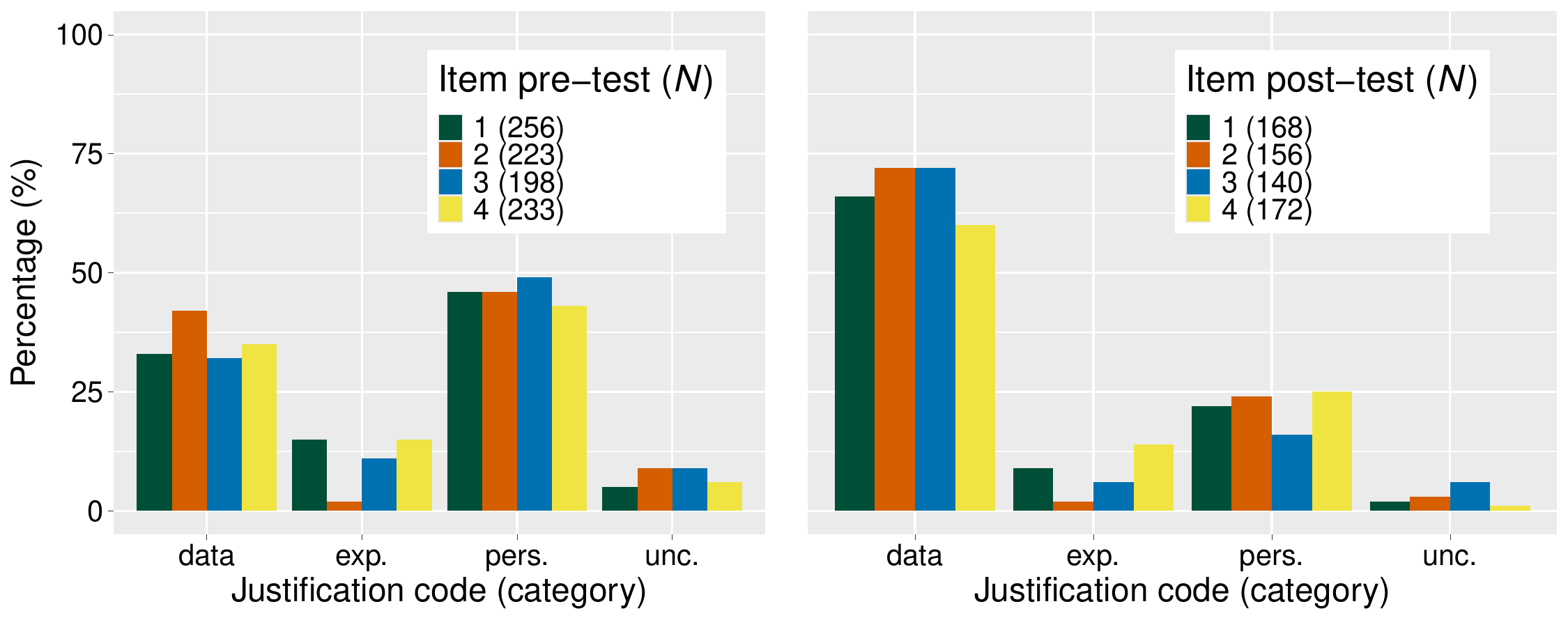}
\caption{Coding of students' justifications in the categories \textit{data}, \textit{exp.}, \textit{pers.}, and \textit{unc.}. Pre-test data are shown in the left graph, post-test results are reported in the right graph. Each item of the questionnaire is indicated with a different color as shown in the figure label. The numbers $N$ indicated in the parenthesis indicates the total number of codes assigned for each item in the pre- and post-test respectively.}
\label{fig:justifications_separateitems_allinstruments}
\end{figure*}

To investigate whether the findings discussed so far depend on the different questionnaire items, we analyzed the coding per-category of each item separately (see Fig.\ \ref{fig:justifications_separateitems_allinstruments}). 

In both the pre- and post-test (left and right side of Fig.\ \ref{fig:justifications_separateitems_allinstruments} respectively), we observe for each isolated item the tendencies already discussed in the cumulative percentages of Fig.\ \ref{fig:justifications_all} for students to use more often justifications based on data quality and personal reasons and to shift in the post-test to data quality based justifications at the expense of personal reasons. The differences between items for both pre- and post-tests are not prominent with the exception of the category \textit{exp.}\ for item 2. 
In fact, for this item students choose the category \textit{exp.}\ for this item much less frequently than for the other three items of the questionnaire. This finding can be interpreted considering that the instrument choices for item 2 are instruments that are very similar in function and form. They only differ by the reading type.

The difference between items in the pre-test distributions is significant with a small-sized effect [$\chi^2(9)=31.64$, $p<0.001$, $w=0.19$].

In the post-test, using simulated $p$-values, we again obtain a significant, small-sized effect [$\chi^2(9) = 29.44$, $p<0.001$, $w=0.22$]. 

Since these small statistical differences are clearly due to the category \textit{exp.}\ of item 2, as discussed above, we conclude that students' reasoning is mainly consistent in different contexts, i.e., for different items.

\begin{figure*}[!]
\includegraphics[width=.7\linewidth]{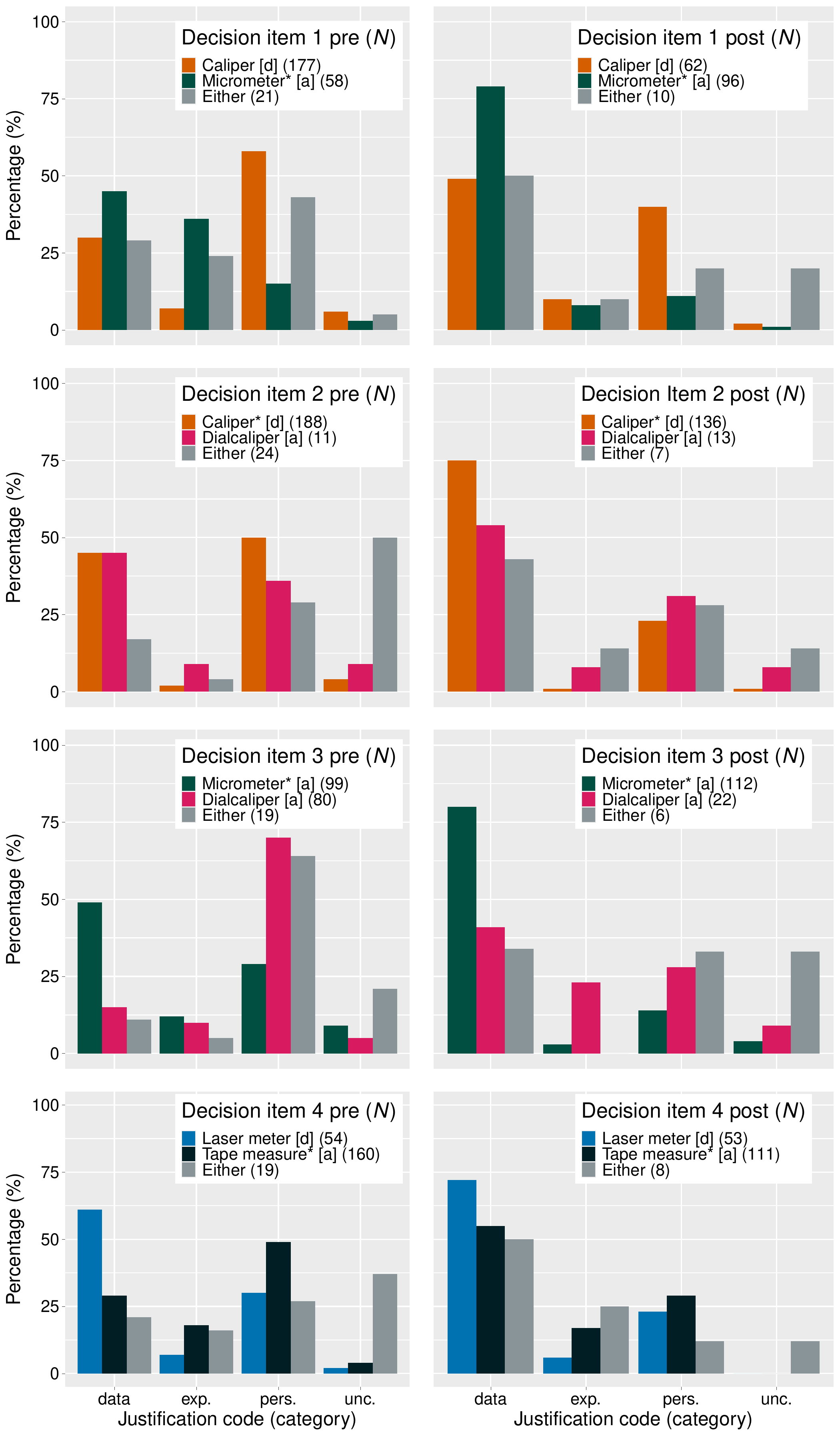}
\caption{Percentage of students' justifications in the four coding categories as a function of instrument chosen in the multiple choice part of the questionnaire. Top to bottom graphs show respectively items 1 to 4. Results for the pre-test are on the left side of the figure, while post-test results are on the right side of the figure. In the legend, the type of instrument (digital or analog) is indicated for each instrument as $[d]$ and $[a]$ respectively. The percentage here is calculated using the number $N$ of justifications used for each instrument, which is indicated in the parenthesis. Note that the sum of the those percentages for each type of answers sum up to \SI{100}{\percent}. Instruments with lower uncertainty are indicated with an asterisk.}
\label{fig:justificationsperinstrument}
\end{figure*}

To better understand the dependence on instrumentation of reasoning, we finally considered students' reasoning behind their decisions about a specific instrument and plotted in Fig.\ \ref{fig:justificationsperinstrument} the percentage of codes for each category, each item and instrument separately. On the left side of the figure, we plot the pre-test results, while on the right side the post-test results. 
For clarity, we indicated in the figure that the instruments are digital or analog with the symbols "[d]" and "[a]" next to the instrument name. 

This analysis allows us to investigate the reasoning behind the choice of a specific instrument. We describe here the results for each item separately, while leave for the discussion section the comparison between items.

For item 1 we observe that in the pre-test students who chose the digital caliper tend to use justifications based on personal experience, easiness of use and personal biases (category \textit{pers.}), while when they opt for the micrometer, they consider more often data quality (category \textit{data}) and practical consideration specific to the instrument procedure (category \textit{exp.}). The choices "Either of the two" (labeled “Either” in the figure) are mainly due to personal reasons in the pre-test and to data quality reasons in the post test.

After instruction (which involved the use of micrometers), the majority of justifications used for both instruments' choice are related to data quality considerations, in accordance to the results of analysis discussed above. However, a certain group of students persisted in using justifications based on personal considerations when choosing the digital caliper.

It is interesting that the small sub-group of students that selected the digital caliper and justified their choice with data quality reasons in the post-test is arguing against the technical information provided to students regarding the measurement uncertainties.

We found for item 1 a statistical significant difference of medium effect between the justifications of the decisions options in both pre-test [$\chi^2(6) = 44.76 $, $p<0.0001$, $w=0.42$] and post-test [$\chi^2(6) = 33.73 $, $p<0.001$, $w=0.45$].

These results can be interpreted considering the fact that the micrometer has, in fact, the smaller measurement uncertainty and that probably most students have less experience and confidentiality with micrometers with respect to calipers.

Note that for the statistical analysis we just emphasis item 1 because of the higher number of justifications for each instrument and the more spread distribution  of decisions.

For item 2, for both instruments (digital caliper and analog dial caliper), the majority of justifications are coded in the categories \textit{data} and \textit{pers.}. 
We do not observe a marked difference between the reasoning behind the choice of the two instruments. This can be understood considering that the two instruments are very similar in form and usage. Also for this item, for both instruments, the majority of the justifications in the post-test are based on data quality considerations and only a small amount on personal considerations.

The choices "Either of the two" use mainly justifications that are uncodable or due to personal reasons in the pre-test, while justifications based on data quality and personal reasons in the post test.

For item 3, in the pre-test students motivate the choice of the micrometer because of data quality and personal considerations, while the choice of the analog dial caliper mainly for personal considerations. This situation changes in the post-test in which justifications for the micrometer are based almost entirely on data quality reasons and justifications for the analog dial caliper are almost evenly distributed between the first three categories.
When students chose the option "Either of the two" their justifications are due to personal reasons in the pre-test, while justifications are evenly distributed between the categories \textit{data}, \textit{pers.}, and \textit{unc.}\ in the post-test.

Finally, considering item 4 we found that in both pre- and post-test students justify their choice for the laser meter mainly based on data quality considerations. For the analog measurement tape, students more often use personal considerations in the pre-test, while considerations based on data quality in the post test. The justifications for the option "Either of the two" are assigned almost evenly to all four categories in the pre-test, while are more often in the category \textit{data} in the post-test.

\section{\label{sec:Discussion}Discussion}

The analysis of students' decisions of instrument (see Fig.~\ref{fig:decisions}), indicates that, when the majority of students initially selected the instrument with larger measurement uncertainty, the distribution of choices changes after instruction and students tend to choose instruments with lower uncertainty in the post-test. Conversely, when the majority of students had already selected the instrument with smaller uncertainty before instruction, they generally maintained their choice afterward. 
This pattern aligns with the analysis of students’ justifications, which revealed a post-instruction shift toward reasoning based on data quality. Nevertheless (see Fig.~\ref{fig:justificationsperinstrument}), a small proportion of students continued, after instruction, to inconsistently justify their decisions using data quality based statements while still selecting the instrument with higher uncertainty.

Moreover, by analyzing students' decisions, we found (see Fig.~\ref{fig:decisions}) that in both the pre- and post-questionnaires, more than \SI{90}{\percent} of participants selected one of the two available instruments, whereas fewer than \SI{10}{\percent} opted for the alternative “Either of the two” (labeled “Either” in the figure). The underlying reason for this pronounced preference remains uncertain.
This pattern might reflect an intrinsic tendency among students to commit to a specific choice, even in the absence of an explicit measurement objective that would justify one instrument over another. On the other hand, it could indicate a perceived expectation to select a single “correct” answer, a norm often reinforced in educational contexts. While this observation is intriguing, additional research is required to better understand the cognitive or contextual factors influencing such decision behavior.

Considering students' justifications for their choice of instruments (see Fig.~\ref{fig:justifications_all}) we found that students decisions are based on seven different aspects that we referred above as sub-categories. Although these sub-categories cannot be used for statistical analysis, they capture the full range of nuances in students’ justifications, providing a comprehensive picture of the factors that influence students decision-making. Students' justifications show several perspectives including aspects related to the measurement itself, but also on the personal experience and preferences. When scaffolding lab teaching, instructors can consider students' perspective and guide their work based on the results of this categorization. To our knowledge, there is no study that creates a manual to characterize students' thinking about instruments in physics laboratory courses. 
In fact, in the study of Logman et al.\ \cite{logman2025} students’ decisions were (deductively) characterized according to Svenson’s decision-making framework \cite{Svenson} and did not inductively explore students' reasoning. 

In our study, we also found that while before instruction students base their decisions mainly on personal reasons, after instruction, the majority of justifications regard data quality. This can be a result of students having acquired experience in using the instruments during the lab course, but also because of their learning on measurement uncertainties and systematic effects. 

Similarly to our findings, Logman et al.\ \cite{logman2025} reported that prior experience with an apparatus influenced students’ choices. However, in their study, only about \SI{20}{\percent} of students based their decisions on technical considerations (classified as level 3 decision making). If we regard our \textit{data} category as corresponding to the third level of decision making in Logman et al.\ work, we observe a substantially higher proportion of such reasoning among our participants. We believe that this difference might be due to the fact that our study involved simpler instrumentation as opposed to more advanced lab equipment (e.g., oscilloscopes). Furthermore, differences in the timing of the studies in which the course took place and the different number of laboratory sessions in which students participated may also have had an impact. 

Looking at the differences between questionnaire items (Fig.\ \ref{fig:justifications_separateitems_allinstruments}), we conclude that the results discussed so far are consistent for the different contexts of the different questionnaire's items. The only apparent difference is for item 2, for which the category \textit{exp.}\ occurs (for both pre- and post-test) much more rarely with respect to the other items. This is understandable considering the fact that in item 2 students have to decide between two very similar instruments in working mechanism, usage and in shape, differing only on their reading way.

The analysis of students' justification by instrument (see Fig.\ \ref{fig:justificationsperinstrument}), let us conclude that, independently on the instrument, students shift their justifications from personal considerations towards data quality considerations upon instruction. When they use justifications regarding data quality, they are mostly in agreement with the choice of the instruments with lower measurement uncertainties.

For the two questionnaire items with the micrometer (items 1 and 3, see Fig.\ \ref{fig:justificationsperinstrument}), we observe that this instrument is chosen before instruction less often by students. They chose the other instrument in both cases mainly for personal reasons. This can be interpreted as a result of students not being familiar with micrometers.

In both items 1 and 2, it seems like students have a strong preference at the pre-survey towards the digital instrument. In item 4 however, this is not the case. Further analysis of students justifications did not indicate that students have a bias towards analog or digital instrumentation. In fact we find some justifications that show a bias (category \textit{pers.3}), but this category does not occur very often and in this case, the bias can be in both ways, either thinking that digital is better or vice-versa. Moreover, as discussed previously students use justifications based on data quality and choose independently of analog or digital instrument the one with lower uncertainty.

Before concluding, we discuss the limitations of this study. First, we find that it is not possible to assume that students use terminology regarding measurement uncertainties and systematic effects correctly. In fact, it is known from the literature that students' struggle to understand several aspects of measurement uncertainties \cite{PhysRevSTPER.4.010108, GeschwindMesUnc, kok2023c, hull2020}. This makes the findings of this study limited, as we have to consider several aspects of data quality reasoning together and cannot clearly distinguish when students decide for an instrument based on its precision, accuracy or correctness of the measurement result.

Secondly, this study only considers very simple instruments and very simple measurement tasks. Although this choice is suitable for investigating students' basic reasoning when choosing instruments, we are aware that the results are limited to simple cases. 

\section{\label{sec:Conclusion}Conclusion}

In this study, we have investigated students' decisions about simple measurement instrumentation before and after instruction, and we have inductively developed a coding manual that allows us to characterize students’ reasoning when selecting those instruments. The coding manual indicates that students consider in their reasoning data quality, experience, ease of use, practicability but also personal biases and fun.
We have found that laboratory instruction influenced both students’ decisions and their justifications, leading to a stronger emphasis on data quality. In fact, after instruction, the majority of students choose the instrument with lower uncertainty and base their justifications mainly on data quality reasons and less often than before instruction on personal experience and intuition. 
These findings suggest that dedicating special laboratory sessions on measurements and their uncertainty can change students' habits in the laboratory and encourage them to base their choices on evidence rather than intuition. However, we highlight the importance of discussing with students the tight relationship between the goal of a measurement and the decision for an instrument. This is important to develop "expert-like" attitudes in the laboratory and avoid that students focus on performing measurement as precise as possible regardless of the situation and when it is not needed. In fact M.A. plans in the future to include as a part of the course a discussion and activity about this topic.

No clear evidence was found of a systematic bias among students toward digital or analog instruments. When such a bias exists, it appears to regard only a very small proportion of students. For this sub-group, an explicit discussion of potential biases of digital versus analog instrumentation is recommended. However, future research should examine the existence of such a bias in greater detail. The present study can be used as a foundation for developing a multiple-choice questionnaire specifically designed to investigate this aspect.

We finally conclude that laboratory instruction on measurements may be enriched by encouraging students to actively choose their measuring instruments and to reflect on the rationale for their choices. This approach can be used for both informing instructors about students’ reasoning when selecting instruments and for helping students develop the habit of explicitly reflecting on their decision-making processes.

\section{Acknowledgments}
M.A. and E.P. would like to thank Dr. Michael F. J. Fox for organizing an International Teaching Lab meeting at the Department of Physics, Imperial College London, in April 2022. This study started as a discussion at \emph{The Gloucester Arms} in South Kensington.

\bibliography{bibliography}

\afterpage{\onecolumngrid
\appendix*
\section{Coding Manual}
Here is the coding manual as it was used in this study:

{\renewcommand{\arraystretch}{1.5}
\renewcommand{\tabcolsep}{0.15cm}
\begin{table}[b]
\footnotesize
\caption{\label{tab:coding_manual_full}The coding manual}
\begin{tabular}{p{.07\textwidth} p{.2\textwidth}p{.2\textwidth}p{.45\textwidth}}

\hline\hline
Code & Definition & Coding rule & Example\\
\hline
data.1 &
Data quality resulting from the measurement &
The precision and/or accuracy of the instrument or the correctness of the measurement result are considered. &

''For more precise measurements, I would always use the micrometer, as it exceeds the accuracy of calipers.''\newline
''The dial caliper is made of plastic and can be deformed by mechanical action. This makes the measurement inaccurate.''\newline
''It also seems more reliable, as the accuracy of roll-up measuring tapes diminishes over time.''\\
\hline
data.2 & 
Avoid mistakes, systematic errors or measurement uncertainties by the reading &
Instruments that show a reading prevent for uncertainties or systematic errors that can happen when taking a measurement reading & 

''Human reading errors are minimized.''\newline
''Reading errors can be minimized with the digital laser device.''\\

\hline
exp.1 &
Practical considerations specific to the measurement &
The instrument is considered more or less suitable/appropriate to measure the quantity needed because of the instrument functioning or the properties of the physical quantity/object to be measured & 
''Because it covers the area in which measurements are to be taken. ''\newline
''Laser does not work because no surface is perpendicular to the table.''\newline
''Lasers require an end point at which the reflection can be detected.''\newline
''Because you can hook it onto the edges of the table.''\newline
''a) looks like it would be better suited for round rods.''\newline
''Since it looks like you could easily place the metal rod there and thus measure the diameter well.''\\

\hline
pers.1 & 
Personal experience &
The instrument is (not) unknown by the student, or it is clearer for the student what the instrument does, or the student used the instrument already in the past. & 
''Device A is what I know and have used before. The other device is just unknown.'' 

''I knew how to use that device, but I don't know the other one.''\\

\hline
pers.2 &
Practical personal considerations specific to the instrument usage &
The instrument is considered easier, more comfortable, or faster to use or read.  & 
''Easier to read. ''\newline
''Is easier to use. ''\newline
''The tape measure is easier to handle than the laser (which requires batteries and is also more expensive to buy; you always have a tape measure lying around). ''\newline
''Because the laser measuring device is faster.''\newline
''Measuring tape is easier to use.''\\

\hline
pers.3 &
Personal bias  &
Students consider one instrument to be more modern or better than the other, this can based on intuition or gutfeeling. & 
''Looks more modern.''\newline
''Is safer.''\newline
''The measuring device gives me more confidence that it is correct.''\newline
''Classic and old-fashioned.''\newline
''More trusted''\\

\hline
pers.4 &
Fun, Enjoyment &
Students motivate their choices based on fun in using a tool or on the desire to use it & 
''\dots but analog is more fun.''\newline
''It's more fun because of the addition.''\newline
''More likable''\\

\hline
unc.1 &
Uncodable & Students uses reasons that are not clearly understandable or do not make sense. &
''I would insert the metal rod between the calipers and then read the value.'' 
\newline
''I can't make up my mind.''\\
\hline\hline
\end{tabular}
\end{table}
}
\twocolumngrid
}

\end{document}